\documentclass[namedreferences]{solarphysics}
\usepackage[hyperref,optionalrh,solaromanenum]{spr-sola-addons} 
\usepackage{graphicx}                    
\usepackage{color}                       
\usepackage{breakurl}                         
\usepackage{epstopdf}
\usepackage{overpic}

\usepackage{comment}

\begin{document}

\begin{article}

\begin{opening}

\title{Delving into the Historical Ca {\sc II} K Archive from the Kodaikanal Observatory: the Potential of the Most Recent Digitised Series}

%
\author[addressref={1,2},corref,email={chatzistergos@mps.mpg.de, theodosios.chatzistergos@inaf.it}\orcid{https://orcid.org/0000-0002-0335-9831}]{\inits{T. }\fnm{Theodosios }\lnm{Chatzistergos}}
\author[addressref={1}]{\inits{I.}\fnm{Ilaria~Ermolli}}
\author[addressref={2,3}]{\inits{S. K. }\fnm{Sami~K.~Solanki}}
\author[addressref={2}]{\inits{N. A. }\fnm{Natalie~A.~Krivova}}
\author[addressref={4}]{\inits{D. }\fnm{Dipankar~Banerjee}}
\author[addressref={4}]{\inits{B. K. }\fnm{Bibhuti~K.~Jha}}
\author[addressref={4}]{\inits{S. }\fnm{Subhamoy~Chatterjee}}

%
\runningauthor{Chatzistergos et al.}
\runningtitle{Potential of Kodaikanal Ca {\sc II} K series}

\address[id={1}]{INAF Osservatorio Astronomico di Roma, Via Frascati 33, 00078 Monte Porzio Catone, Italy}
\address[id={2}]{Max Planck Institute for Solar System Research, Justus-von-Liebig-weg 3,	37077 G\"{o}ttingen, Germany}
\address[id={3}]{School of Space Research, Kyung Hee University, Yongin, Gyeonggi 446-701, Republic of Korea}
\address[id={4}]{Indian Institute of Astrophysics, Koramangala, Bangalore 560034, India}

\begin{abstract}
Full-disc Ca {\sc II} K photographic observations of the Sun carry direct
information about the evolution of solar-plage regions for more than a
century and are therefore a unique dataset for solar-activity
studies. For a long time Ca {\sc II} K observations were barely explored,
but recent digitisations of multiple archives have allowed their
extensive analysis. However, various studies have reported diverse
results partly due to the insufficient quality of the digitised
data. Furthermore, inhomogeneities have been identified within the
individual archives, which, at least partly, could be due to the
digitisation.  As a result, some of the archives, \textit{e.g.} that from the
Kodaikanal observatory, were re-digitised. The results obtained by different 
authors who analysed the data from the new digitisation of the Kodaikanal archive
differ from each other as well as from those derived from the old digitisation. 
Since the data were processed and analysed using
different techniques, it is not clear, however, whether the
differences are due to the digitisation or the processing of the
data. To understand the reasons for such discrepancies, we
analyse here the data from the two most recent digitisations of this
archive. We use the same
techniques to consistently process the images from both archives and
to derive the plage areas from them. Some issues have been identified
in both digitisations, implying that they are intrinsic
characteristics of the data. Moreover, errors in timing of the observations
plague both digitisations. Overall, the most recent 16-bit
digitisation offers an improvement over the earlier 8-bit one. It also
includes considerably more data and should be preferred.
\end{abstract}

%
\keywords{Solar Cycle, Observations; Instrumental Effects}

\end{opening}

%
\section{Introduction}
Full-disc photographs of the Sun in the resonance K line of the singly-ionized calcium, Ca II, at 3933.67 \,\AA~ were first obtained in the early 1890's by Henri Alexandre Deslandres and George Ellery Hale \citep{hale_solar_1893} with the spectroheliographs developed at the Paris and Kenwood observatories, respectively. 
Since then, regular observations with similar instruments have been performed  at various sites around the globe, \textit{e.g.} at the Kodaikanal (since 1904), Mt. Wilson (since 1915), Mitaka (since 1917), Coimbra (since 1926), and Arcetri (since 1931) observatories. 
Ca~{\sc II}~K observations with interference filters started later, e.g at the Rome (since 1964), Kandilli (since 1968), and Big Bear (since 1981) observatories. 
These observations are one of the main sources of information on the long-term changes in the lower solar chromosphere, that is the first thousand kilometres above the temperature minimum. 
Furthermore, the Ca~{\sc II}~K line provides information about solar magnetism. 
This is due to the large increase in the intensity of the line core when sampling bright magnetic regions \citep[plage; see \textit{e.g.}][]{skumanich_sun_1984}. 
The potential of the full-disc Ca~{\sc II}~K observations to serve as a proxy of the magnetic field \citep[][and references therein]{schrijver_relations_1989,loukitcheva_relationship_2009,chatzistergos_recovering_2019} makes them very valuable for studies of the evolution of solar activity, as well as for analyses of the magnetic activity of stars other than the Sun. 
Also, irradiance reconstructions and Earth's climate-variability studies can significantly benefit from analyses of Ca~{\sc II}~K observations. 

\begin{figure}
	\centering
\includegraphics[width=1\linewidth]{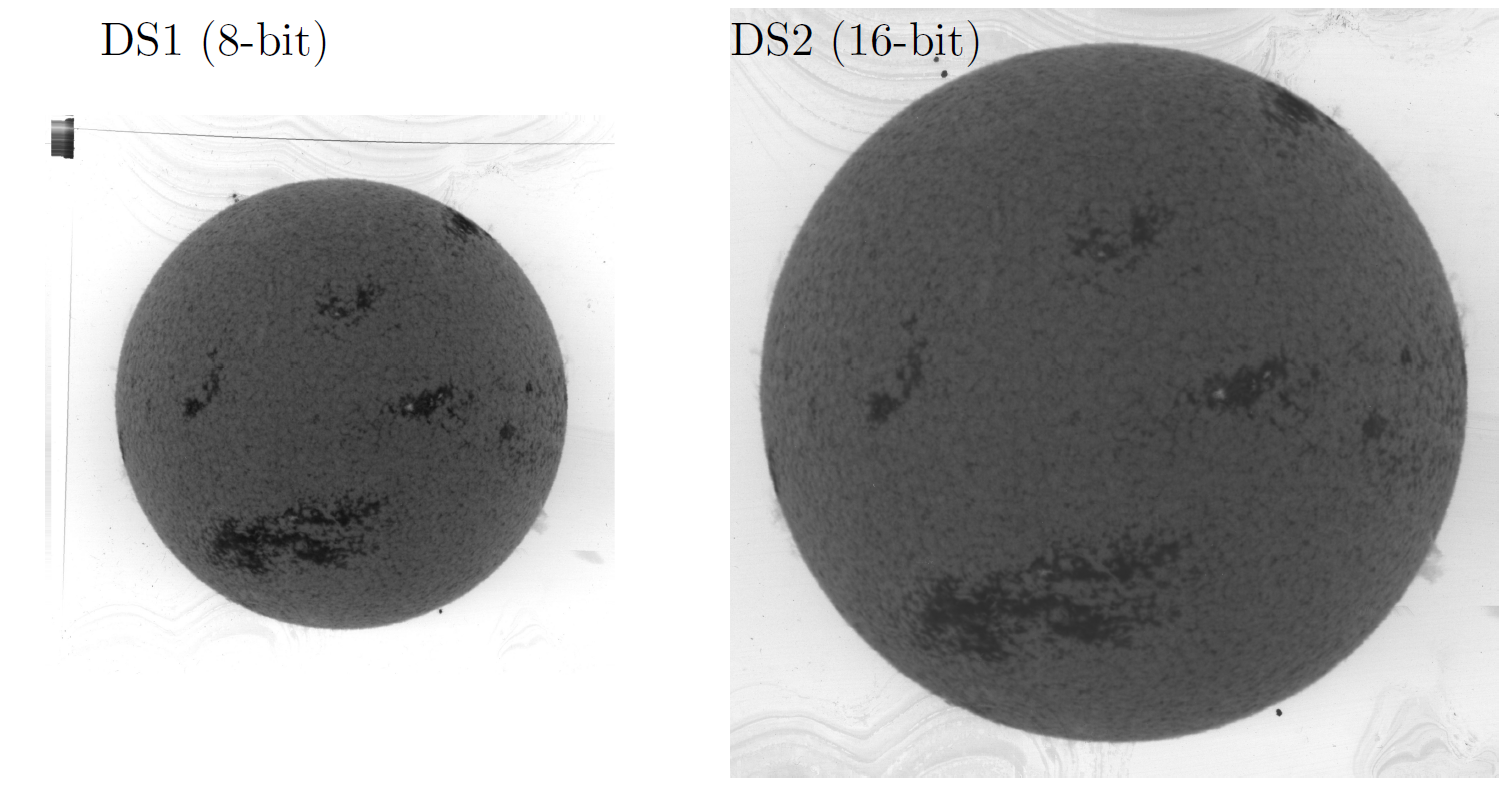}
	\caption{Examples of the images from the two digitisations of the photographic Ca {\sc II} K observations of the Kodaikanal observatory taken on 02 January 1936. Left: 8-bit digitisation from data set 1, or DS1 in short; Right: 16-bit digitisation from DS2. 
		The images are shown to their full range of values and were not compensated for the ephemeris. The DS1 image is shown to scale with the DS2 image (i.e. equal number of pixels per cm), while the DS2 has been cropped to a width of $2R+R/6$, where $R$ is the solar radius in pixels.}
	\label{fig:rawimgexample_19360102}
\end{figure}

\begin{figure}
	\centering
	\includegraphics[width=1\linewidth]{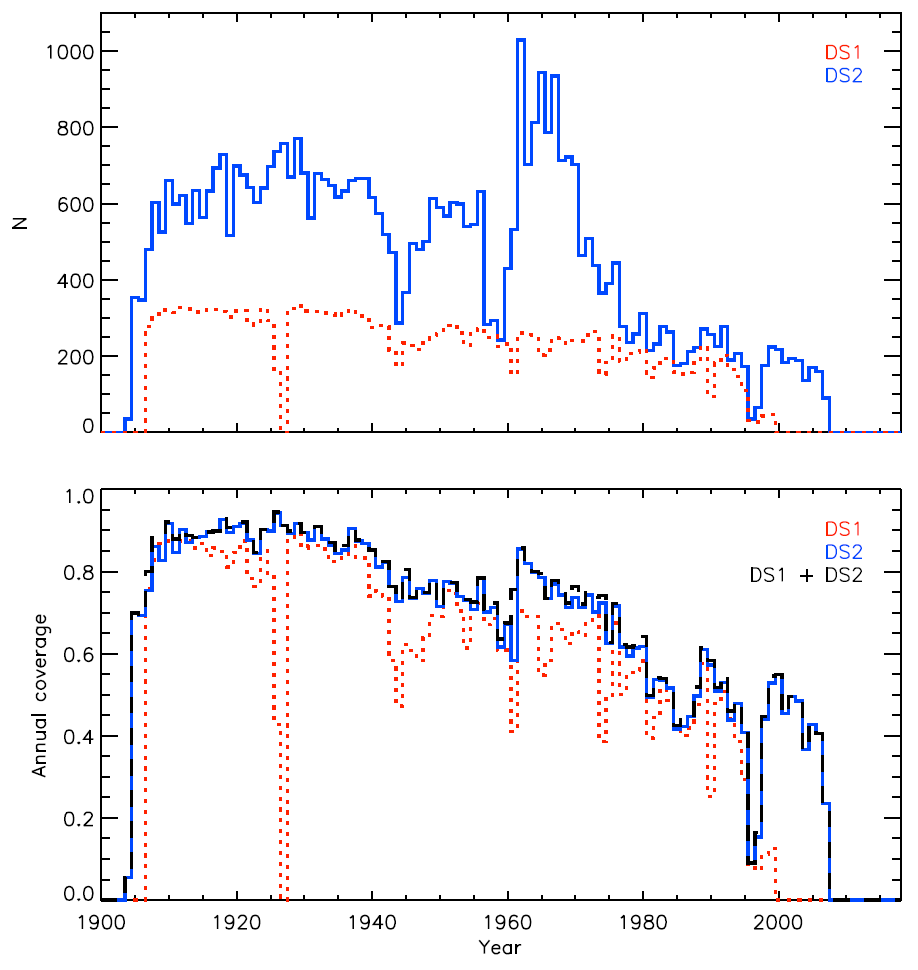}
	\caption{Number of images per year (\textit{top panel}) and annual coverage (\textit{bottom panel}) of the Kodaikanal data in the two digitised series. DS1 is shown in dotted red, DS2 in solid blue, while the annual coverage by the two archives together is shown in dashed black.}
	\label{fig:ndataused}
\end{figure}

Most of the Ca {\sc II} K observations performed so far are stored in photographic archives, a number of which have recently been digitised \citep[see][]{chatzistergos_analysis_2017,chatzistergos_analysis_2019,chatzistergos_historical_2019}.
For example, the photographic archives from the Arcetri (\,1931--\,1974),  Kodaikanal (\,1904--\,2007), Kyoto (\,1926--\,1969), McMath-Hulbert (\,1948--\,1979), Meudon (\,1893--\,2002), Mitaka (\,1917--\,1974), Mt. Wilson (\,1915--\,1985), Rome (\,1964--\,1979), and Sacramento Peak (\,1960--\,2002) observatories have been made available in digital form. 
The availability of the Ca~{\sc II}~K series as digital data has initiated their extensive exploitation for studies of the long-term variation of the chromospheric magnetic field and for a variety of retrospective analyses. 
For example, \cite{foukal_behavior_1996},   \cite{ermolli_comparison_2009}, \cite{tlatov_new_2009}, \cite{chatterjee_butterfly_2016}, and \cite{priyal_long-term_2017} presented plage-area time series derived from the analysis of different archives with distinct image-processing methods. 
\cite{harvey_cyclic_1992}, \cite{ermolli_digitized_2009}, and \cite{chatterjee_butterfly_2016} produced butterfly diagrams of plage regions from the Mt. Wilson, Arcetri, and Kodaikanal observations, respectively. 
Moreover, \cite{sheeley_carrington_2011}, \cite{chatterjee_butterfly_2016}, and \cite{pevtsov_reconstructing_2016} produced Carrington maps from Ca~{\sc II}~K observations of  the Mt. Wilson and Kodaikanal archives. 
Recently, \cite{chatterjee_variation_2017} analysed the supergranulation scale variation in historical Kodaikanal Ca {\sc II} K data following a previous similar analysis of modern data by \cite{ermolli_measure_2003} with Rome Precision Solar Photometric Telescope (Rome/PSPT, hereafter) observations. 

\begin{figure}
	\centering 
	\includegraphics[width=1\linewidth]{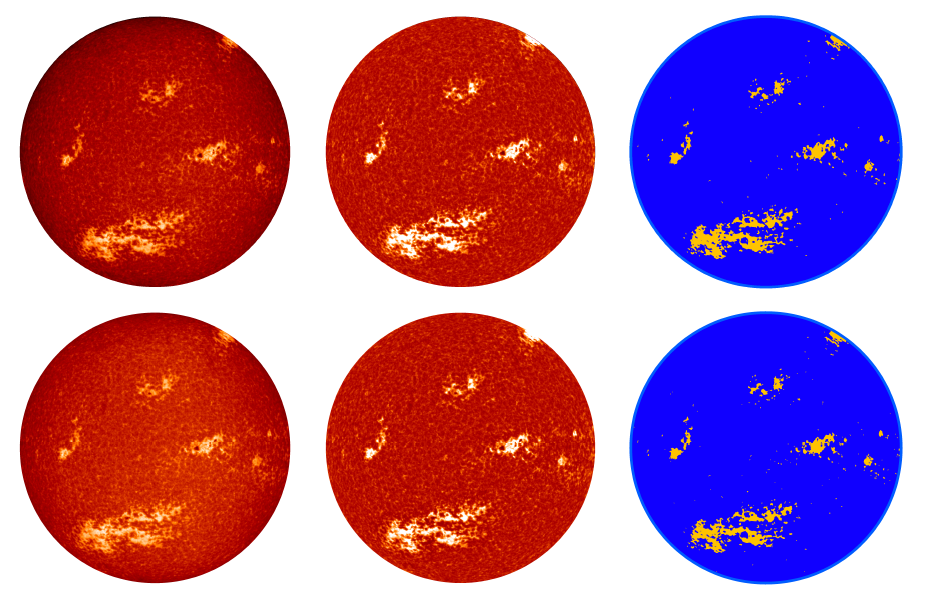}
	\caption{Selected processing steps applied to images from the two digitisations of Kodaikanal observations taken on 02 January 1936 and shown in Figure \ref{fig:rawimgexample_19360102}. The DS2 (16-bit) image is shown in the \textit{upper row}, while the DS1 (8-bit) one in the \textit{lower row}. The columns show the original density image, photometrically calibrated and limb darkening corrected image, and segmentation mask, respectively. The raw images are shown over the entire range of values found within the solar disc, while the limb-darkening-compensated images are saturated at contrast values of [-1,1]. The segmentation masks show the QS regions in blue and the plage regions in orange.}
	\label{fig:processedimagessamedayflat} \end{figure}

\begin{figure*}
	\centering
	\includegraphics[width=1\linewidth]{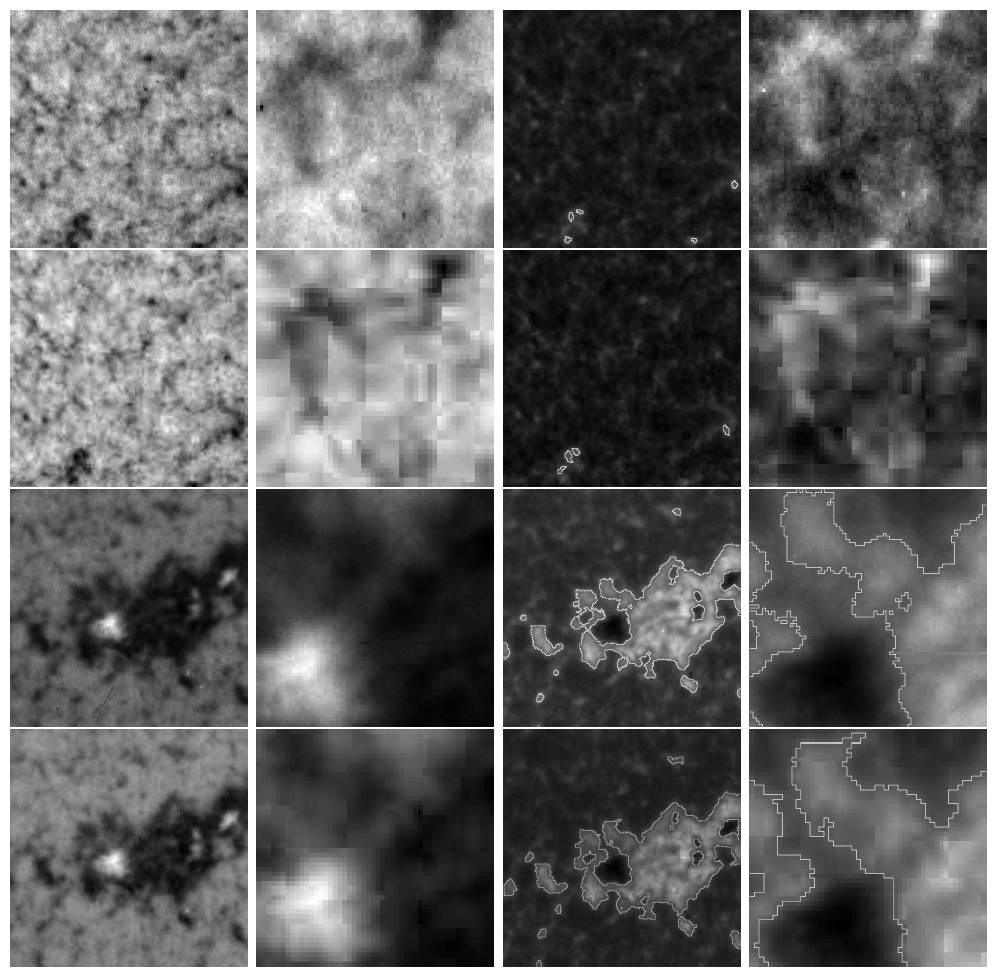}
	\caption{Magnified sections from the raw DS1 and DS2 negative images (\textit{first two columns}) and calibrated and limb-darkening compensated images (\textit{last two columns}) of the Kodaikanal observation taken on 02 January 1936 (see Figures \ref{fig:rawimgexample_19360102} and \ref{fig:processedimagessamedayflat}) displaying a quiet Sun region (\textit{top 2 rows}) and a plage region with a sunspot (\textit{bottom 2 rows}). The image from DS2 is shown in \textit{rows 1 and 3}, while the one from DS1 is shown in \textit{rows 2 and 4}. The sections in \textit{columns 1+3} and \textit{2+4} have widths of $260''$ and $65''$, respectively, which correspond to 200 and 50 pixels for the DS1 image, respectively. The images are shown over their full range of values. \textit{Over-plotted contours} show the plage regions identified with the method applied in our study.}
	\label{fig:rawimgexample_zoomin_193601020}
\end{figure*}

\begin{figure}
	\centering
	\includegraphics[width=1\linewidth]{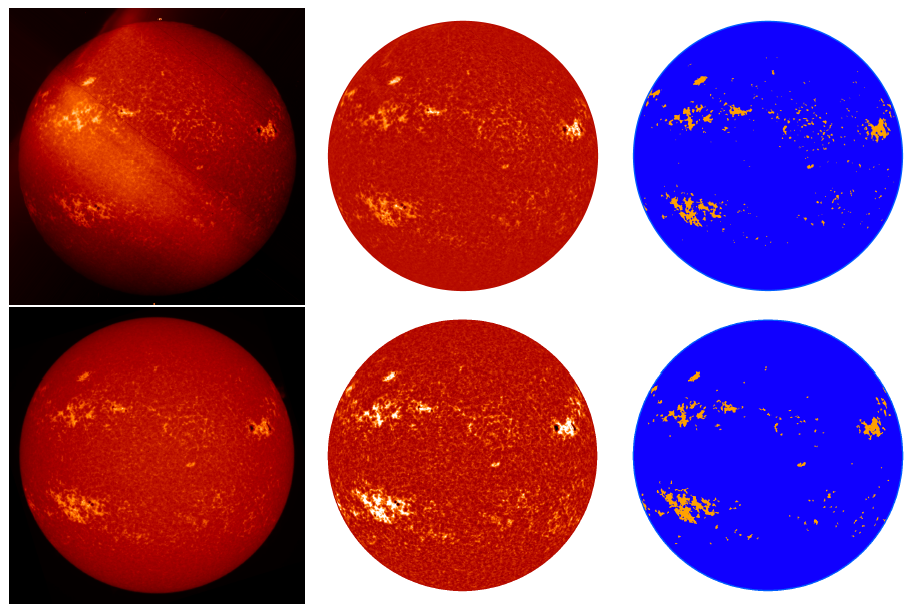}
	\caption{Comparison of observations from DS2 (\textit{upper row}) and Rome/PSPT (\textit{lower row}), both taken on 27 May 2000. Shown are the raw images (\textit{left column}), processed and limb-darkening compensated images (\textit{middle column}), and the segmentation masks (\textit{right column}). The raw images are shown over the entire range of values found within the solar disc, while the limb-darkening compensated images are saturated at [-0.5,0.5]. In the segmentation masks the QS regions are coloured blue and the plage regions orange.}
	\label{fig:comparisonpspt}
\end{figure}

\begin{figure}
	\centering
	\includegraphics[width=1\linewidth]{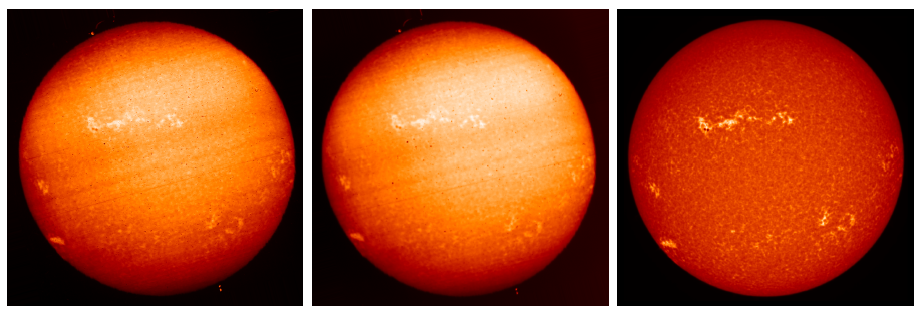}
	\caption{Raw density images from DS1 (\textit{left}), DS2 (\textit{middle}), and Rome/PSPT (\textit{right}) taken on 03 April 1999 illustrating saturated plage regions in the Kodaikanal data.}
	\label{fig:comparisonpsptsatregions}
\end{figure}

\begin{figure}
	\centering
	\includegraphics[width=1\linewidth]{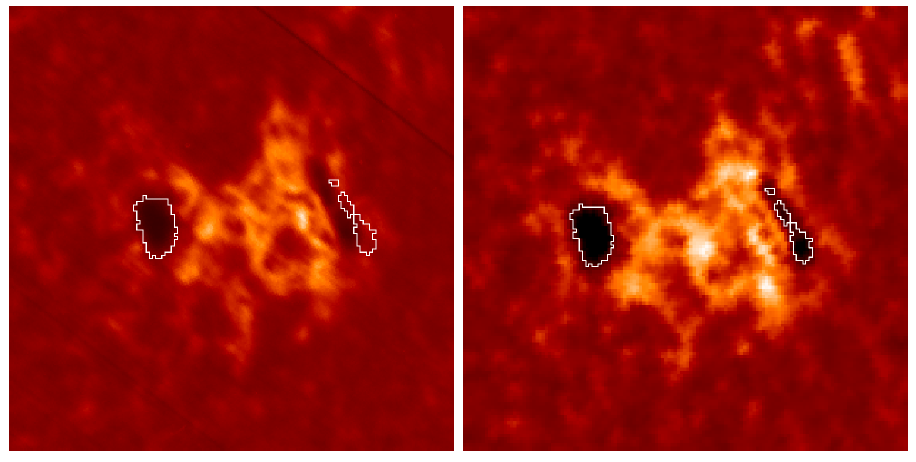}
	\caption{Enlargement of observations displayed in Figure \ref{fig:comparisonpspt} from DS2 (\textit{left}) and Rome/PSPT (\textit{right}) showing an active region. The images are saturated at [-0.5,0.9]. The contours mark the location of sunspots as identified in the Rome/PSPT image. See Section \ref{sec:imagecomparison} for more details.}
	\label{fig:200027-5-0-0kodaplagezoomin}
\end{figure}

Most studies performed on historical Ca~{\sc II}~K data confirm  some well-known characteristics of past cycles, \textit{e.g.} they all report that Solar Cycle 19 showed remarkably high plage coverage and a broad latitudinal distribution of active regions. They also agree on the overall increase of solar activity over the first half of the 20th century and the decrease over the last decades.
However, they also show significant inconsistencies, such as large differences in absolute values and in the detected short- and long-term trends \citep{ermolli_potential_2018,chatzistergos_analysis_2019}. 
Such differences are seen both, between the results derived from different archives, and between the same archives but processed and analysed with different methodologies.

Some of these discrepancies can be ascribed to the varying quality of the analysed data partly arising from the digitisation. 
Most of the series were digitised independently at different periods and using various set-ups. 
Furthermore, some series underwent multiple digitisations, \textit{e.g.} the Kodaikanal, Meudon, Mitaka, and Mt. Wilson archives. 
The two main reasons for the re-digitisation of the various series were: i) the availability of higher quality digitising devices than those employed in the older digitisations and ii) various problems identified in some of the earlier digitised series.

Among the photographic archives of Ca {\sc II} K observations, the one from the Kodaikanal observatory has probably the largest collection of images, covering more than a century with nearly daily observations  between 1904 and 2007. It is thus a particularly important archive. 
The Kodaikanal photographic archive underwent three digitisations over the last three decades \citep{kariyappa_variability_1994,makarov_22-years_2004,priyal_long_2014}, although only the last two digitisations included the entire solar disc. 
In \cite{chatzistergos_analysis_2019} we consistently analysed with the same method the data from the former digitisation of the Kodaikanal archive along with the archives from Arcetri, McMath-Hulbert, Meudon, Mitaka, Mt. Wilson, Schauinsland, and Wendelstein to derive plage areas. We used the Kodaikanal and Mt. Wilson series as reference to construct two composites of plage areas over the entire 20th century. The two composites showed different absolute levels and discrepancies over different Solar Cycles. To understand whether some of these differences are coming from the digitisation and can potentially be resolved with the new digitisation of the Kodaikanal data and to best exploit the potential of one of the most important Ca~{\sc II}~K series, we here analyse the quality and intrinsic differences in the images derived from the two most recent digitizations of the full-disc Ca~{\sc II}~K observations from the Kodaikanal observatory.
We process the data from both digitisations consistently to derive plage areas and to compare the results to each other, as well as to those presented in the literature.

This article is structured as follows: 
in Section \ref{sec:data} we give an overview of the data employed in this study and describe the methods used to process them. 
We compare the images from the two digitisations in Section \ref{sec:characteristicsdatasets} where we also discuss characteristics of the series.
We present our results of plage areas from the last digitisation of the Kodaikanal data in Section \ref{sec:plageareas} and compare them to other series reported in the literature. 
Finally we draw our conclusions in Section \ref{sec:conclusions}.

\section{Data and Processing}
\label{sec:data}
\subsection{Data}
We analyse the data from the two most recent digitisations of the photographic full-disc Ca~{\sc II}~K Kodaikanal observations. 
These data were taken with a spectroheliograph having a nominal bandwidth of 0.5\,\AA.
The first dataset (DS1, hereafter) was obtained by  \cite{makarov_22-years_2004} by scanning 22,158 photographic observations taken from 1907 to 1999 with a linear array of 900 pixels. 
The data were stored in JPG format as 1800$\times$1800 pixel$^2$ images, with 8-bit accuracy, and an average pixel-scale of $1.3''$pixel$^{-1}$. 
The second dataset (DS2, hereafter) was derived by  \cite{priyal_long_2014} by scanning 48,928 photographic observations taken from 1904 to 2007 with a CCD camera with the resolution of 4096$\times$4096 pixel$^2$. 
The data were stored in FITS format with 16-bit accuracy and an average pixel scale of $0.9''$pixel$^{-1}$.
Figure \ref{fig:rawimgexample_19360102} shows the two digitisations of a single Kodaikanal photographic plate taken on 02 January, 1936. 
The raw negative images are displayed to their full range of values, while the image from the DS1 is shown to scale to the DS2 one showing equal number of pixels per cm.

Figure \ref{fig:ndataused} shows the number of images per year in DS1 and DS2. 
Only a sub-sample of the available photographic plates were scanned for DS1, while almost all of the existing plates have been scanned for DS2.
We find only 13,835 images of each set referring to the same solar observations, being scans of the same photographic plate. 
The fraction of days within each year with at least one observation from either Kodaikanal series is also shown in Figure \ref{fig:ndataused}. 
We find 21,405 days with at least one observation from both archives (out of 21,746 and 26,492 days in DS1 and DS2, respectively), as demonstrated by the mismatch between the coverage by DS2 and the total coverage by DS1 and DS2 together. Besides, we notice that there are seven years (1907, 1909, 1911, 1913, 1955, 1973, and 1993) over which DS1 has a better coverage than DS2. 
These discrepancies can, at least partly, be explained by errors in reporting both the date and timing of the observation in the digitised files.
Such errors in the digitised series are not surprising considering the large number of plates and should most likely affect images from both digitisations. We find that the errors in the dates are limited to only a few images, while the errors in  the time of the observations affect a considerable amount of data. In this regard, it is worth noting that images from DS2 include almost the entire plate, thus allowing us to compare the date on the plate to the one passed in the meta-data of the file. This is not the case for the images from DS1, which have been cropped to include the solar disc and only a small area outside of the disc.
Furthermore, there is an inconsistency in both datasets in the time format. Time is given as Indian Standard Time up to the 1960's and as Coordinated Universal Time afterwards. For our study, all times from both datasets were corrected to be in Indian Standard Time. Note that this choice does not affect the results presented in the following, since the analysis was done on daily mean values.

The radius of the solar disc varies with time in both series, but it has an average value of 685 and 1095 pixel for DS1 and DS2, respectively. 
A heliostat was employed at the Kodaikanal Observatory, which allows determining the orientation of the solar disc based on the date and time of the observation.
However, the orientation of the plates during the digitisation introduced another orientation that needs to be taken into account to orient the images. This angle is usually small and in general random. 
Markings have been introduced on the plates before the scanning \citep{priyal_long_2014} to identify the north and south poles of the solar disc.
However, the information about which side corresponds to the east or west is lost.
The information from the regions of the original plates lying outside the solar disc seems to have been maintained in DS2. 
In DS1, however, there are images where these regions have been saturated (\textit{e.g.} Figure \ref{fig:rawimgexample_19360102}).
Various artefacts, such as scratches and emulsion holes, appear in both digitisations. 
Many other artefacts, such as dust or  hairs, are found at different locations in images of the two digitisations.
One should also note that the DS2 series was generated more than a decade after the DS1 one, so natural degradation of the photographic plates is also partly responsible for this discrepancy. 
Moreover, since the 8-bit digitisation was performed with a linear array, this caused a few images to have rows that are offset in the $x-$direction and hence have distorted discs. 
This issue is resolved in the 16-bit digitisation, where a CCD camera was used.

Both sets of digitised images require instrumental calibration of the digitisation camera before any further processing. 
Such a calibration has not been applied to the DS1 \citep{ermolli_comparison_2009} data, but it has partially been applied to the DS2 data and the standard data for instrumental calibrations were stored. 
In particular, the dark current was removed with a built-in program of the scanning device, while a lab-sphere illuminating a white surface was used to measure the flat-field of the CCD. 
The same surface was used to support the photographic plates during their scanning. 
Inspection of the available data shows that this surface does not always cover the same area as the plate. As a result, information outside the disc is lost when dividing the digitised observation by the corresponding flat-field image. 
Furthermore, the exposure time of the flat was not constant and was different from the one used for scanning the plates. 
Since the same CCD recorded the image with different gain in four quadrants, small errors in the photometric calibration of the data could affect the calibrated images showing variations over the quadrants and other residual inhomogeneities. In addition, flat-field images are expected to change over the course of the digitisation, and it is important to have such images created close to the scanning time of each image.
However, flat-field images are missing from several folders in the archive or the included flat image does not always account completely for the difference in the quadrants. 

We also use modern CCD-based observations taken with the Rome/PSPT as a comparison because there is partial overlap with both digitised Kodaikanal series.
PSPT is located at Monte Porzio Catone and is operated by the INAF Osservatorio Astronomico di Roma \citep{ermolli_prototype_1998,ermolli_photometric_2007}.
Observations with Rome/PSPT started in May 1996 and continue to the present.
The observations used here are taken with an interference filter centred at the Ca {\sc II} K line with bandwidth of 2.5\,\AA. 
The images have dimensions of 2048$\times 2048$ pixel$^2$ and are stored in FITS files after the standard instrumental calibration \citep[][]{ermolli_photometric_2007}.
Therefore, such data can be used to investigate the differences between photographic and CCD observations, as well as between observations taken with a spectroheliograph and an intereference filter.

\subsection{Methods}
In our study we analysed raw DS1 images without calibration of the digitising device, raw DS2 data divided by the flat image taken closest in time to the scanning time, and calibrated Rome/PSPT data.
Some key characteristics of these archives are listed in Table \ref{tab:characteristiscs}.

We used the DS1 and Rome/PSPT data already processed with the technique described by \cite{chatzistergos_analysis_2018,chatzistergos_ca_2018,chatzistergos_analysis_2019} and in the present study we applied exactly the same processing to the data from DS2.
In brief, the images were converted to density images and then photometrically calibrated with a calibration curve (CC, hereafter). The CC was derived by relating the centre-to-limb variation (CLV, hereafter) measured in quiet-Sun regions (QS, hereafter) on the historical observations to a standard reference of QS CLV as measured in modern Rome/PSPT Ca {\sc II} K observations, and linearly extrapolated to the non-QS regions. 
Contrast images were constructed by removing the limb darkening, which was determined in an iterative process. 
This includes application of a running-window median filter and polynomial fitting along rows, columns, and radial locations after the bright features had been removed. 
The contrast images were then segmented with a multiplicative factor to the standard deviation of the QS intensity values. This factor was determined with a method based on the approach of \cite{nesme-ribes_fractal_1996}. 
The multiplicative factor for identifying plage was chosen to be $8.5$. 
This choice resulted in plage areas from the Rome/PSPT that are comparable to those from the SRPM segmentation scheme \citep{fontenla_semiempirical_2009} applied to the same Rome/PSPT data. 
We emphasize that for this work the precise value of the segmentation parameter is of no importance, since the main goal here was to process all series consistently to understand some of the difference between the results presented in the literature.
Figure \ref{fig:processedimagessamedayflat} shows examples of the density images and the processed images (calibrated and limb-darkening compensated images) as well as the segmentation masks for the observation shown in Figure \ref{fig:rawimgexample_19360102}.
Images with severe artefacts (\textit{e.g.} missing parts of disc) were excluded from analysis as was done by \cite{chatzistergos_analysis_2018,chatzistergos_analysis_2019}.

\section{Characteristics of Datasets}
\label{sec:characteristicsdatasets}

\subsection{Image Comparison}
\label{sec:imagecomparison}

Here we compare the images from DS1 and DS2 to study the differences due to the various digitisation set-ups. As an example, Figure \ref{fig:processedimagessamedayflat} shows the same solar observations in the two datasets.
The full-disc images appear rather similar; however, a more detailed inspection reveals significant differences.
Figure \ref{fig:rawimgexample_zoomin_193601020} displays enlarged areas of the observation shown in Figure \ref{fig:rawimgexample_19360102} including a QS and a plage region with a sunspot. 
The different magnifications correspond to regions with a width of 200 and 50 pixels, respectively, in the DS1 image.
Due to the lower spatial resolution of the data of DS1 compared to those of DS2, the pixel size in DS1 is slightly larger than in DS2 when projecting the solar disc in the same physical dimensions.
There are evident compression effects in the DS1 images, manifested as smoothed $\approx8\times8$ pixel$^2$ regions. 
The images from DS2 reveal much finer solar structures. 
Furthermore, the effect of compression seems to affect the QS regions in DS1 more than the plage area.

Figure \ref{fig:rawimgexample_zoomin_193601020} shows also the enlarged areas of the calibrated and limb-darkening-compensated images. 
Contours outline the regions that were identified as plage in each image.
The regions identified as plage appear larger and coarser in the DS1 data compared to the DS2 ones.

We now compare the images from DS2 to those from the Rome/PSPT.
Figure \ref{fig:comparisonpspt} shows an image from DS2 and Rome/PSPT before and after selected processing steps. 
The DS2 and Rome/PSPT datasets overlap during the period \,1996--\,2007. 
Over that period, we generally notice saturated regions in the DS2 data, while the QS CLV in DS2 data is stronger than in the Rome/PSPT images. One example of an image with saturated regions in the DS2 data is shown in Figure \ref{fig:comparisonpsptsatregions} along with the image from the DS1 and Rome/PSPT of the same day. The images from both DS1 and DS2 display saturated regions, while the DS2 image is blurred compared to the DS1 one, suggesting that it might have been taken out of focus.
The contrast in the calibrated and CLV-compensated DS2 image is lower than that in the Rome/PSPT one, which is contrary to the expectation due to the narrower bandwidth of the Kodaikanal observations. This suggests that Kodaikanal observations over that period suffer from severe stray-light effects.
The identified plage regions from both images appear very similar, yet with some differences. 
More small regions can be identified in the DS2 image compared to the Rome/PSPT one, while large plage regions appear more extended in the Rome/PSPT image than in the DS2 one. 
We note, however, that the regions appear different to some degree due to the time difference between the observations of DS2 and Rome/PSPT, which were taken with a time difference of about six hours on average, with the Kodaikanal one preceding the Rome/PSPT observation.

Figure \ref{fig:200027-5-0-0kodaplagezoomin} shows a magnified section of a plage region from the processed images displayed in Figure \ref{fig:comparisonpspt}. 
This sub-array was extracted close to the limb and the plage region is smaller but more squeezed in DS2 than in the Rome/PSPT observation. 
We also marked with contours the regions identified as sunspots in the Rome/PSPT image, by picking the regions that lie below $\bar{C}-3\sigma$, where $\bar{C}$ is the mean contrast over the disc and $\sigma$ is the standard deviation of contrast values within the disc. 
The same contours are over-plotted in the DS2 image, scaled using the radius ratio of the two images. 
The sunspot regions cover roughly the same area, despite the difference in nominal bandwidth used by the two observatories (0.5\,\AA~and 2.5\,\AA~for Kodaikanal and Rome/PSPT data, respectively). 
This suggests that the Kodaikanal observations, at least the ones overlapping with Rome/PSPT, might have been taken with a broader bandwidth than the nominal one \citep[see also][]{chatzistergos_analysis_2019}. 
However, these could also be, at least partly, differences due to the use of a filter and a spectroheliograph in the Rome/PSPT and DS2 data, respectively. The exact effect of the bandwidth on the sunspot areas would require further investigation.
Furthermore, the contrast of the sunspot regions in the DS2 is reduced compared to that in the Rome/PSPT images. This might be due to stray-light or underexposure over the very dark regions of the disc.

\subsection{Characteristics of Time Series}
\label{sec:characteristicsseries}

\begin{table*}
	\caption{Characteristics of the analysed archives. See Section \ref{sec:characteristicsseries} for more details.}
	\centering
	\begin{tabular}{lccc}
		\hline
		&  DS1 &  	DS2 &  	Rome/PSPT \\
		\hline 
		Period covered				& \,1907--\,1999	 &\,1904--\,2007	  & \,1996--\,2018\\
		Number of images			&22158			 &48928			  &3292\\
		Number of images used		&19291			 &45519			  &3292\\
		Number of days				&21746			 &26492			  &3287\\
		Number of days used			&18963			 &25781			  &3287\\
		Spectral bandwidth	[\,\AA]	&0.5			 &0.5			  &2.5\\
		Pixel scale	[$''$pixel$^{-1}$]		&1.3			 &0.9			  &2\\
		Disc eccentricity   		&  $0.10\pm0.05$ &  $0.11\pm0.06$ &  $0.05\pm0.02$\\
		Spatial resolution [$''$]	    &  $3.5\pm1.6$   &   $3.3\pm1.2$  &   $5.3\pm0.7$\\
		Inhomogeneities				&  $0.07\pm0.02$ &   $0.07\pm0.03$&   $0.03\pm0.02$\\
		$\sigma_D$ raw images		&  $0.09\pm0.04$ &   $0.23\pm0.06$&   -\\
		$\sigma_C$ processed images	&  $0.06\pm0.02$ &   $0.05\pm0.02$&   $0.06\pm0.02$\\
		$\sigma_C^{\mathrm{QS}}$ processed images&  $0.020\pm0.008$ &   $0.019\pm0.010$ &   $0.021\pm0.003$\\
		Reduced $\chi^2$			&$0.002\pm0.002$ &$0.002\pm0.002$ &   -\\
		\hline
	\end{tabular}
	\label{tab:characteristiscs}
\end{table*}

\begin{figure}
	\centering
	\includegraphics[width=1\linewidth]{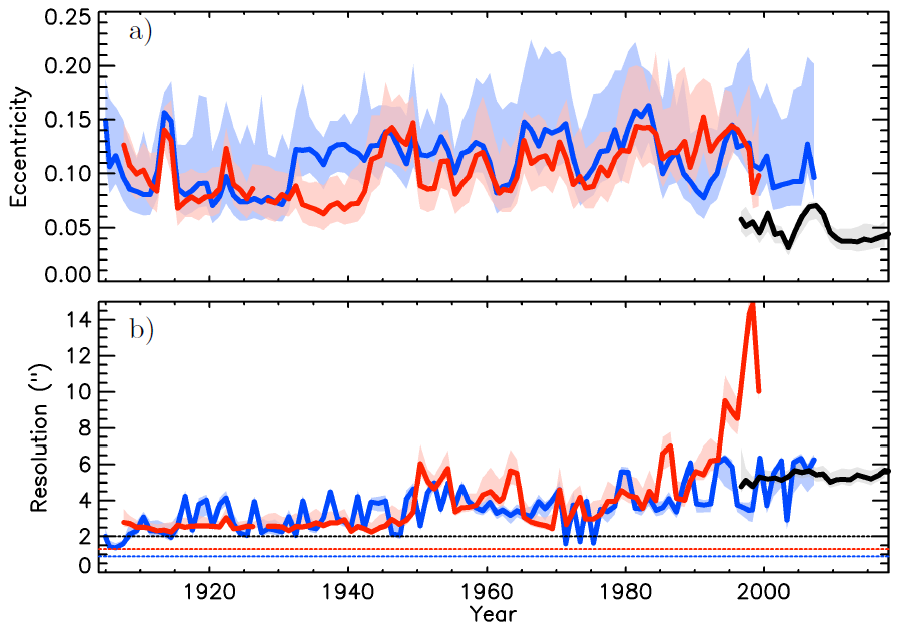}
	\caption{Solar disc eccentricity (\textit{top panel}) and spatial resolution (\textit{bottom panel}) computed for DS1  (red), DS2 (blue), and Rome/PSPT (black) data as a function of time. Shown are annual mean values (solid lines) along with the asymmetric $1\sigma$ interval (shaded surfaces). The horizontal dotted lines in the \textit{lower panel} indicate the values for the average pixel scale of the data in the  archives.}
	\label{fig:eccentricities}
\end{figure}

\begin{figure}
	\centering
	\includegraphics[width=1\linewidth]{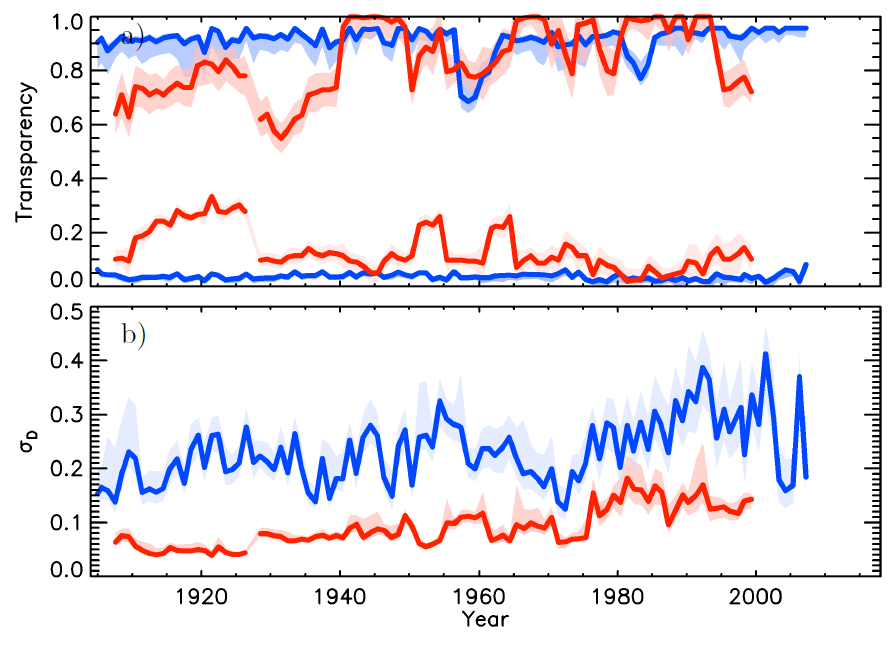}
	\caption{The maximum and the minimum values of transmittance (\textit{top panel}) and the standard deviation of density values, $\sigma_D$, (\textit{bottom panel}) within the solar disc for the DS1 (red) and DS2 (blue) data as a function of time. Shown are annual mean values (solid lines) along with the asymmetric $1\sigma$ interval (shaded surfaces). These quantities are not defined for the Rome/PSPT data, which are given directly in units of intensity.}
	\label{fig:density_raw}
\end{figure}

\begin{figure}
	\centering
	\includegraphics[width=1\linewidth]{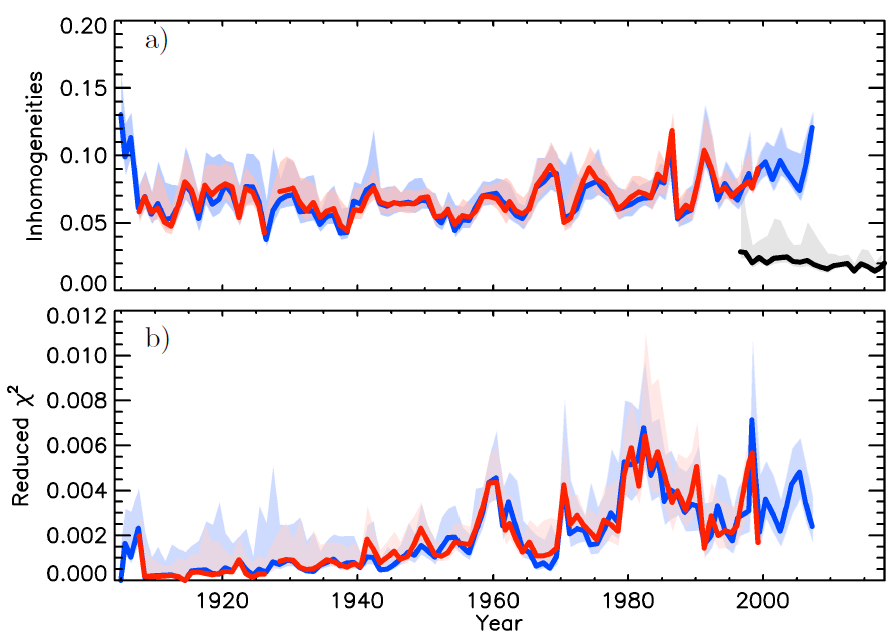}
	\caption{Identified large scale inhomogeneities (\textit{top panel}) and reduced $\chi^2$ of the fit to the curve obtained by relating the measured density QS CLV from the Kodaikanal data to a reference QS CLV from Rome/PSPT data (\textit{bottom panel}) for DS1 (red), DS2 (blue), and Rome/PSPT (black, only in the top panel) data as a function of time. Shown are annual mean values (solid lines) along with the asymmetric $1\sigma$ interval (shaded surfaces). The $\chi^2$  is not applicable to the Rome/PSPT data.}
	\label{fig:inhomogeneities}
\end{figure}

\begin{figure}
	\centering
	\includegraphics[width=1\linewidth]{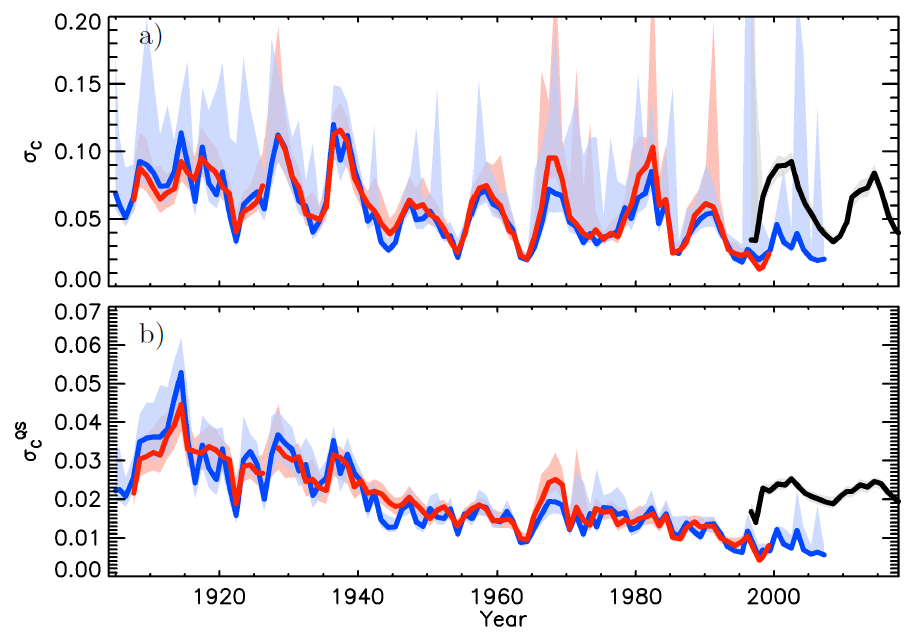}
	\caption{The standard deviation of the contrast values, [$\sigma_C$], over the whole solar disc (\textit{top panel}) and in the quiet Sun only, [$\sigma_C^{\mathrm{QS}}$], (\textit{bottom panel}) for the calibrated and limb-darkening compensated DS1 (red), DS2 (blue), and Rome/PSPT (black) data as a function of time. Shown are annual mean values (solid lines) along with the asymmetric $1\sigma$ interval (shaded surfaces).}
	\label{fig:density_flat}
\end{figure}

To assess the quality of the two Kodaikanal series and to compare it to that of modern data, we consider different image attributes that allow us to describe both global and local properties of the digital data. 
In particular, following \cite{ermolli_comparison_2009} we study the eccentricity of the solar disc, the spatial resolution, the large scale inhomogeneities, similarity of QS CLV to that from the Rome/PSPT, as well as density/intensity characteristics.
Table \ref{tab:characteristiscs} summarizes the results obtained, which are discussed in more detail in the following.

We computed the eccentricity of the recorded solar disc as $e=\sqrt{1-(R_{\mathrm{min}}/R_{\mathrm{max}})}$, where $R_{\mathrm{min}}$ and $R_{\mathrm{max}}$ are the smallest and largest measured radii. 
However, the disc in the historical data does not always have an elliptical shape, rather an irregularly distorted shape due to problems with the drive of the spectroheliograph (affecting data from both DS1 and DS2) or with the digitising device (mainly affecting data from DS1).
It is important to have an estimate of the eccentricity of the solar disc, since \cite{chatzistergos_analysis_2019} demonstrated that the uncertainty in the derived plage areas increases with the disc eccentricity.
The eccentricity of the solar disc from the DS1 and DS2 images is shown in Figure \ref{fig:eccentricities}\textbf{a)}. 
Annual averages are plotted along with the asymmetric 1$\sigma$ interval. 
We get an average value of $\bar{e}=0.10\pm0.05$ for the DS1 and $\bar{e}=0.11\pm0.06$ for the DS2 series.
For both DS1 and DS2, $e$ and $\sigma_e$ increase with time. 
The value of $e$ is relatively low prior to 1940 (DS1) and 1930 (DS2), but it is higher and varying afterwards. 
\cite{ermolli_comparison_2009} had also studied the eccentricity of the disc in the DS1 and reported qualitatively the same results. 
In particular, they also found an increase in $e$ and $\sigma_e$ with time, although the jumps in the results in the early period were less pronounced in their case. 
They found a mean eccentricity of $\bar{e}=0.12\pm0.06$, which is slightly higher than derived here.
The eccentricity we derive for DS1 and DS2 is larger than to that for the Rome/PSPT data, which is on average $\bar{e}=0.05\pm0.02$. 
This value is consistent with the value of $\bar{e}=0.04\pm0.03$ reported by \cite{ermolli_comparison_2009}. The maximum (RMS) error in the plage areas for the average disc eccentricity found for Kodaikanal data is 0.013 (0.0005), while for Rome/PSPT it is 0.002 (0.0003) as reported by \cite{chatzistergos_analysis_2019}.
The eccentricities for the two Kodaikanal series are similar, although the one for DS2 is higher for most of the time, as can be seen from Figure \ref{fig:eccentricities}\textbf{a)}. Figure \ref{fig:eccentricities}\textbf{a)} shows all data, but this is seen also when doing the same plot for the data that have the same date and time. In principle this can be due to a small inclination of the plate when it was digitised, or issues with the code to identify the limb, which might have been affected by artefacts. 

To study the spatial resolution of the images, we evaluate the frequency at which 98\,\% of the power spectral density is taken into account. 
The computation was performed on $64\times64$ disc sub-arrays of quiet-Sun regions. 
We randomly positioned 100 such segments within the inner $R$/3 of the disc, and the average value from all the segments was adopted. 
This method is similar to the approach used by \cite{ermolli_comparison_2009}.
However, we can potentially get a better estimate of the average spatial resolution of the data than \cite{ermolli_comparison_2009}, considering that the solar observation was not recorded instantaneously, but rather in strips with variable spatial resolution.
The derived spatial resolution is shown in Figure \ref{fig:eccentricities}\textbf{b)}. 
We find the resolution for DS1 and DS2 in general to be roughly the same and slowly getting worse with time. 
Around 1980, the spatial resolution of DS1 degrades, sharply reaching an average value of $15''$. 
For the data from DS2, the resolution remains between 3 and $6''$. 
The average values over the whole period are $3.5\pm1.6''$ and $3.3\pm1.2''$ for DS1 and DS2, respectively. 
The spatial resolution of DS2 after 1990 is at similar levels as that of the Rome/PSPT data, but it is considerably worse for DS1 data. This is most likely due to an issue with the digitisation of DS1 data.
The spatial resolution of the Rome/PSPT data is $5.3\pm0.7''$ and roughly constant over the whole period. 
These values are consistent with the $3.3\pm0.1''$ and $5.0\pm0.4''$ for DS1 and Rome/PSPT, respectively reported by \cite{ermolli_comparison_2009}. 

To compare the dynamic range of the data in the two digitisations, Figure \ref{fig:density_raw}\textbf{a)} shows the maximum and minimum values of the raw negative images, in units of transmittance, over the solar disc in the DS1 and DS2 data. 
The values from each dataset were normalised to the maximum value from the respective digitisation.
On average, DS2 is much more stable, with the exception of two periods around 1959 and 1984 during which the maximum values decrease. 
DS1 shows a larger variation with time with abrupt jumps in the transmittance values and gaps. There are periods where the maximum transmittance value is found within the solar disc of DS1 data, hinting at saturation of the low-density regions.
This suggests that DS2 has been digitised in a more consistent manner than DS1.
Figure \ref{fig:density_raw}\textbf{b)} shows the standard deviation of the density values, [$\sigma_D$], over the solar disc from both digitisations. 
The standard deviation in DS2 is consistently higher than that in DS1. 
In both datasets a slight increase of the standard deviation with time is observed, which could be due to an increase in CLV because the employed bandwidth became broader than before or the observations were not centred at the core of the line.

Next, we assess whether and how strongly the images are affected by the large-scale inhomogeneities and artefacts. For this, we compute the relative difference of the image background calculated by the image processing to the QS CLV. 
The image background is a 2D surface map that includes the QS CLV as well as all identified large-scale inhomogeneities and artefacts. This is determined with the iterative process described by \cite{chatzistergos_analysis_2018}.
To make the results between the different datasets comparable, we rescaled (using cubic interpolation) only for this test all images to the same dimensions, such that the radius is always 350 pixel. 
Figure \ref{fig:inhomogeneities}\textbf{a)} shows the values of the relative difference of the background to the QS CLV we get for DS1, DS2, and Rome/PSPT data.
The level of inhomogeneities increases with time for both the DS1 and DS2 data. 
The level of inhomogeneities in the Rome/PSPT decreases with time and at $0.03\pm0.02$ is lower than the values from both Kodaikanal series. 

From the data produced during the image processing, we also analysed the $\chi^2$ of a linear fit to the curve obtained by relating the measured density QS CLV to the logarithm of the reference intensity QS CLV from Rome/PSPT data. 
The reduced $\chi^2$ from this fit can be considered as an indication for instrumental changes, \textit{e.g.} in the bandwidth or central wavelength \cite[see][for a discussion on this]{chatzistergos_analysis_2019}. 
Figure \ref{fig:inhomogeneities}\textbf{b)} shows the derived reduced $\chi^2$ from the fit for images from DS1 and DS2. 
There are no significant differences between the resulting values and their variation with time in the two series. 
The $\chi^2$ of the fit increases with time for both DS1 and DS2, hinting at a gradual degradation of image quality rather than the digitisation.

We have also analysed the contrast values in the calibrated and limb-darkening compensated images. 
Figure \ref{fig:density_flat}\textbf{a)} shows the standard deviation of contrast values over the entire solar disc: $\sigma_C$, for DS1, DS2, and Rome/PSPT.
Unsurprisingly, $\sigma_C$ over the disc qualitatively follows the solar-cycle variability. However, it is less pronounced before 1920, while the amplitude of $\sigma_C$ over Solar Cycles 18, 19, 21, and 22 is lower than the others suggesting some variations in the instrument parameters, or data quality. 
The values from the Rome/PSPT data are at similar levels to those derived for both DS1 and DS2 for most Solar Cycles.
Figure \ref{fig:density_flat}\textbf{b)} shows the same but only for the QS regions: $\sigma_C^{\mathrm{QS}}$.
The standard deviation of the QS decreases with time in both DS1 and DS2 series. 
This can be due to increased underexposure with time or broadening of the bandwidth, if the observation is centred more towards the wing of the line than its core, or due to observational and instrumental effects such as worsening seeing at the observing site, degradation of the spectroheliograph, and changes in the quality of the plates.
The value for the standard deviation of the QS for the Rome/PSPT is always higher than for both Kodaikanal series. This again lends support to the argument that the effective bandwidth of the Kodaikanal over the overlapping period might be broader than the one of the Rome/PSPT or that the Kodaikanal observations were off-centre over that period. 
However, we note that this is not a conclusive test since any of the instrumental or observational issues mentioned above can also contribute to the change of the standard deviation of the QS with time.

Overall, we find both DS1 and DS2 series to exhibit almost the same characteristics and temporal behaviour, suggesting that the studied characteristics are intrinsic to the original Kodaikanal data and are not artefacts of the digitisation.
These include the worsening of the spatial resolution, the increase of the disc eccentricity, the enhancement of the large-scale inhomogeneities, and the change of the CLV with time.
Hence, these should be ascribed to the original photographic observations and not to image issues introduced by the digitisation.
DS2 shows a more consistent distribution of transparency values, suggesting that the digitisation was performed more consistently than for DS1. The data after 1990 show an improved spatial resolution in DS2 data compared to those in DS1.

\section{Plage Areas}
\label{sec:plageareas}
\subsection{Results from DS1, DS2, and Rome/PSPT}
\begin{figure}
	\centering
	\includegraphics[width=1\linewidth]{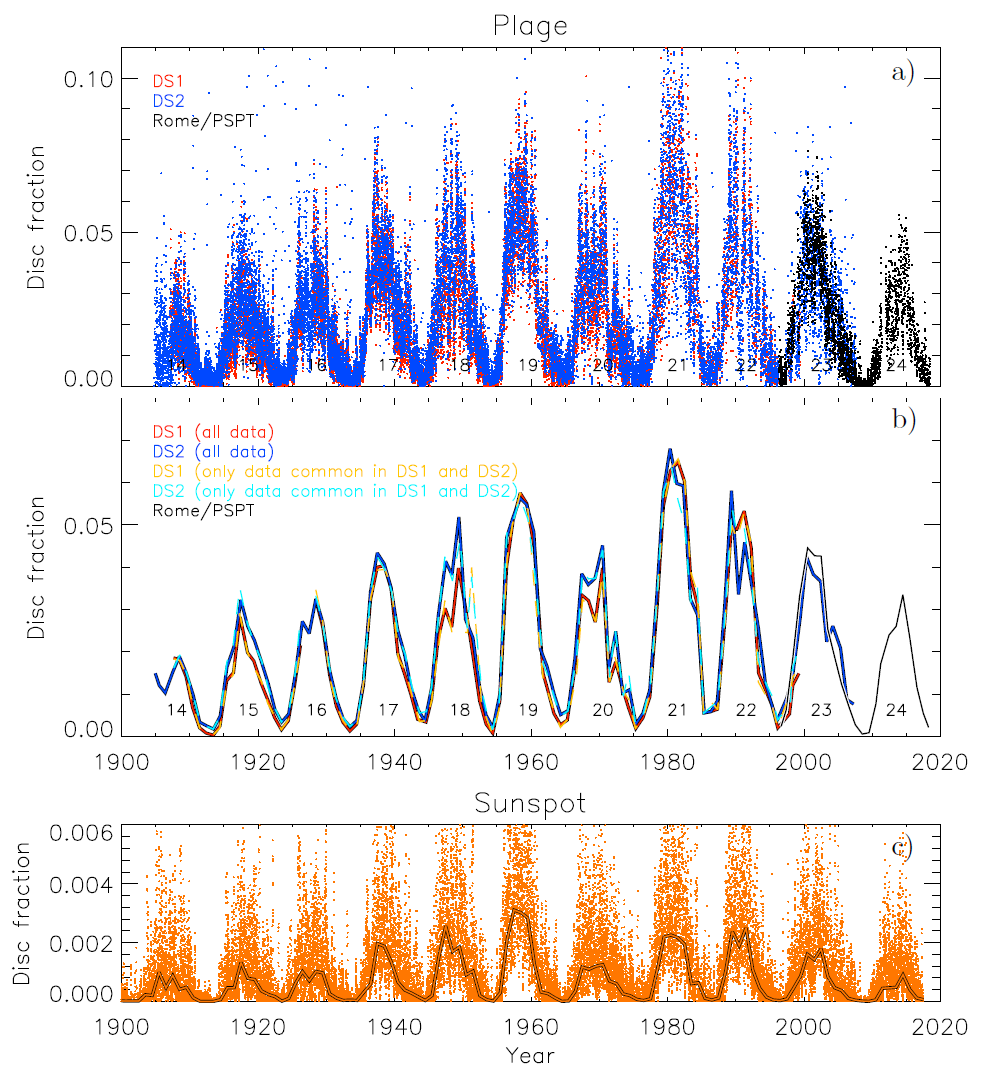}
	\caption{Fractional disc coverage by plage (Panels \textbf{a} and \textbf{b}) as a function of time, derived with the same processing and segmentation parameters from DS1 (red), DS2 (blue), and Rome/PSPT (black) images. Panel \textbf{a} shows daily values, while panel \textbf{b} displays annual mean values (solid lines). The dashed lines in panel \textbf{b} show the annual values for DS1 (orange) and DS2 (light blue) when only the common days in DS1 and DS2 are considered. Also shown (Panel \textbf{c}) are daily (dots) and annual (solid line) values of sunspot areas from \cite{balmaceda_homogeneous_2009}. }
	\label{fig:1discfractionplage}
\end{figure}

We have derived plage areas from DS1 and DS2 processed in the same way.
The results are shown in Figure \ref{fig:1discfractionplage} along with plage areas from the Rome/PSPT data and the sunspot areas by \cite{balmaceda_homogeneous_2009}\footnote{The series has been extended to 31 May 2017 and is available at \url{www2.mps.mpg.de/projects/sun-climate/data.html}}.
We give a tentative RMS error in the derived plage areas from the Kodaikanal and Rome/PSPT data of 0.0025 and 0.0003, respectively, in fraction of the disc area. The value for Kodaikanal data is the sum of the RMS errors due to the disc ellipticity and errors of the processing of the images to perform the photometric calibration as evaluated by \cite{chatzistergos_analysis_2018,chatzistergos_analysis_2019}. 
The value for Rome/PSPT data is acquired by considering only the effect of the disc ellipticity. 
We note, however, that this is not a strictly defined, formal error for the derived plage areas, but our best estimate based on our analysis of synthetic data with our method.

The derived plage areas for DS2 data are distinctly higher than those from DS1 for Solar Cycles 15, 18, and 20 and slightly larger in Cycle 17, while in Cycles 21 and 22 the opposite is the case. 
The two series have RMS difference of 0.006 and Pearson coefficient of 0.95 when daily values are considered (shown in Table \ref{tab:agreement}). 
However, when considering only the observations that have been taken at the same day and same time the RMS difference becomes 0.005 and the Pearson coefficient 0.98.
In Figure \ref{fig:1discfractionplage} we also show the annual plage areas for the common days in DS1 and DS2. 
The RMS differences between the annual values for all data and only the common days in the two series are 0.0004 and 0.003 for DS1 and DS2, respectively. 
The smaller effect on the results for DS1 comes as no surprise considering that it is the series with the fewer observations. 
The agreement between the areas from the two datasets becomes worse for Solar Cycles 20 and 21 when considering only data taken on the same day and time, while it slightly improved for Cycle 22 and remained unchanged for all other cycles. 
Hence, the differences in the computed plage areas do not stem from differences in sampling of the original photographs.
We note, however, that potential errors in the dating of the images could affect our results.
Comparing DS1 and DS2 to the sunspot areas we find that Solar Cycles 21 and 22 in the plage areas fall differ from expectation, being too high relative to the previous cycles. Solar Cycle 18 in DS2 is between Cycle 17 and 19 which is in agreement with the ranking of the Solar Cycles in the sunspot areas.
This is not the case for Solar Cycle 18 in DS1.

\begin{figure}
	\centering
	\includegraphics[width=1\linewidth]{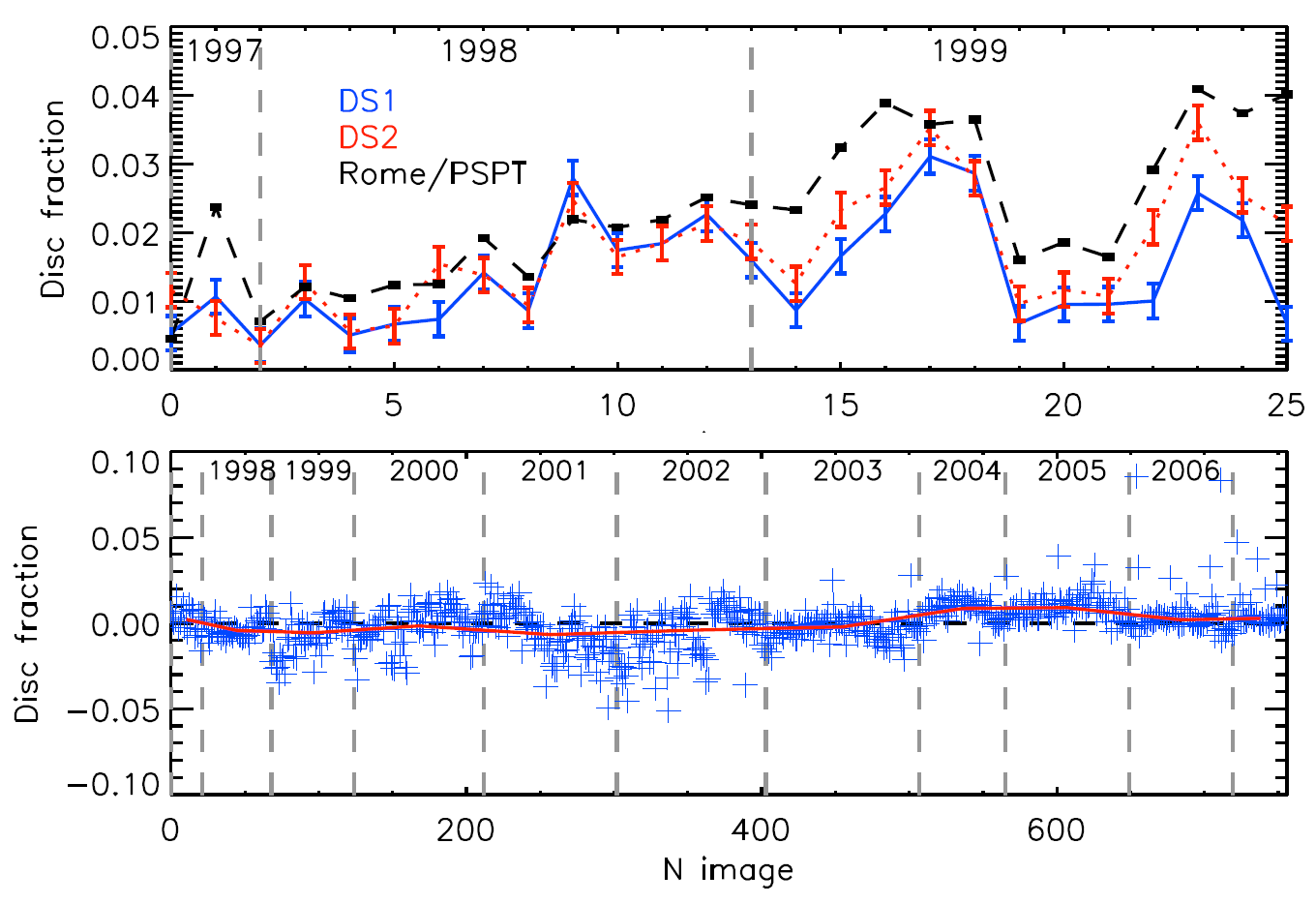}
\caption{\textit{Top: }Plage areas in disc fraction for 26 images taken on the same days found in DS1 (dotted red), DS2 (solid blue), and the Rome/PSPT (dashed black) series plotted against the number of the image. The error bars denote the RMS error in the derived plage areas due to the disc ellipticity and the processing to photometrically calibrate the images (the latter is applicable only to the Kodaikanal data) as found by \cite{chatzistergos_analysis_2019}. \textit{Bottom: } Difference of plage areas derived from the 757 images common to DS2 and Rome/PSPT series plotted against the number of the image. Blue plus signs denote individual values, while the red solid line is for the annual mean value. The dashed vertical lines separate the years, which are written at the top of each panel.}
\label{fig:dfcommondayspspt}
\end{figure}

We notice that the results for Rome/PSPT are in relatively good agreement with those from both the DS1 and DS2 series within the period of overlap. 
The areas from DS2 over the maximum of Solar Cycle 23 are slightly lower by 0.003 than that from the Rome/PSPT, while we also notice an increase in the plage areas of DS2 in 2004, something not seen in the areas from Rome/PSPT or the sunspot areas.
We found merely 26 days with data from all three datasets, DS1, DS2, and the Rome/PSPT in the period 19 May 1997 to 29 May 1999.
Figure \ref{fig:dfcommondayspspt} shows the areas in disc fractions derived from all three datasets, by considering only the days available in all three archives.
For that period, the areas from both DS1 and DS2 lie mostly below the one from Rome/PSPT except for two and four days for which DS1 and DS2 give greater plage areas, respectively.
The RMS difference between the areas derived from DS1 and DS2 to Rome/PSPT is 0.011 and 0.008, respectively, while the maximum absolute difference is 0.033 and 0.019, respectively.

Restricting the comparison between DS2 and Rome/PSPT gives 757 days of overlap over the period 23 April 1997 to 10 September 2007.
The difference between the plage areas derived from DS2 to that from Rome/PSPT is shown in the lower panel of Figure \ref{fig:dfcommondayspspt}. 
We get an average RMS difference of 0.01, while the maximum absolute difference reaches up to 0.08. 
We notice an annual variation in the differences between the series, with Rome/PSPT giving higher plage areas during Winter periods than DS2, while the opposite occurs during the Summer.

\begin{figure*}[t]
	\centering
	\includegraphics[width=1\linewidth]{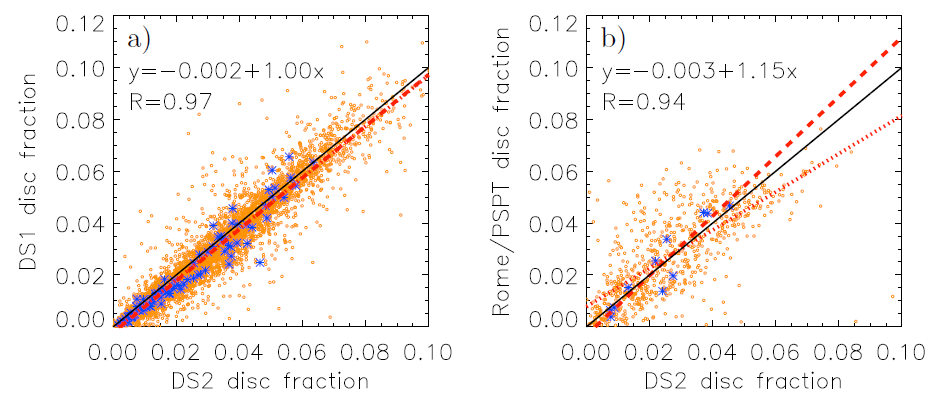}
	\caption{Scatter plots between the plage area values derived from images of DS2 ($x$-axis) and those from images ($y$-axis) of DS1 (Panel \textbf{a}) and Rome/PSPT (Panel \textbf{b}).  Blue asterisks (orange dots) show the annual (daily) values. The solid black lines have a slope of unity and represent the expectation value. The dashed (dotted) red lines are linear fits to the annual (daily) data. Also shown are the corresponding parameters of the linear fits to the annual values and the linear correlation coefficients of the annual values.}
	\label{fig:scatterplotspspt}
\end{figure*}

Figure \ref{fig:scatterplotspspt} shows scatter plots between the derived plage areas from DS2 and those from DS1 and Rome/PSPT. 
We find a good agreement between all series, with linear correlation of 0.97 and 0.94 for DS1 and Rome/PSPT, respectively.

\subsection{Comparison to Other Results}

\begin{figure*}[t]
	\centering
	\includegraphics[width=1\linewidth]{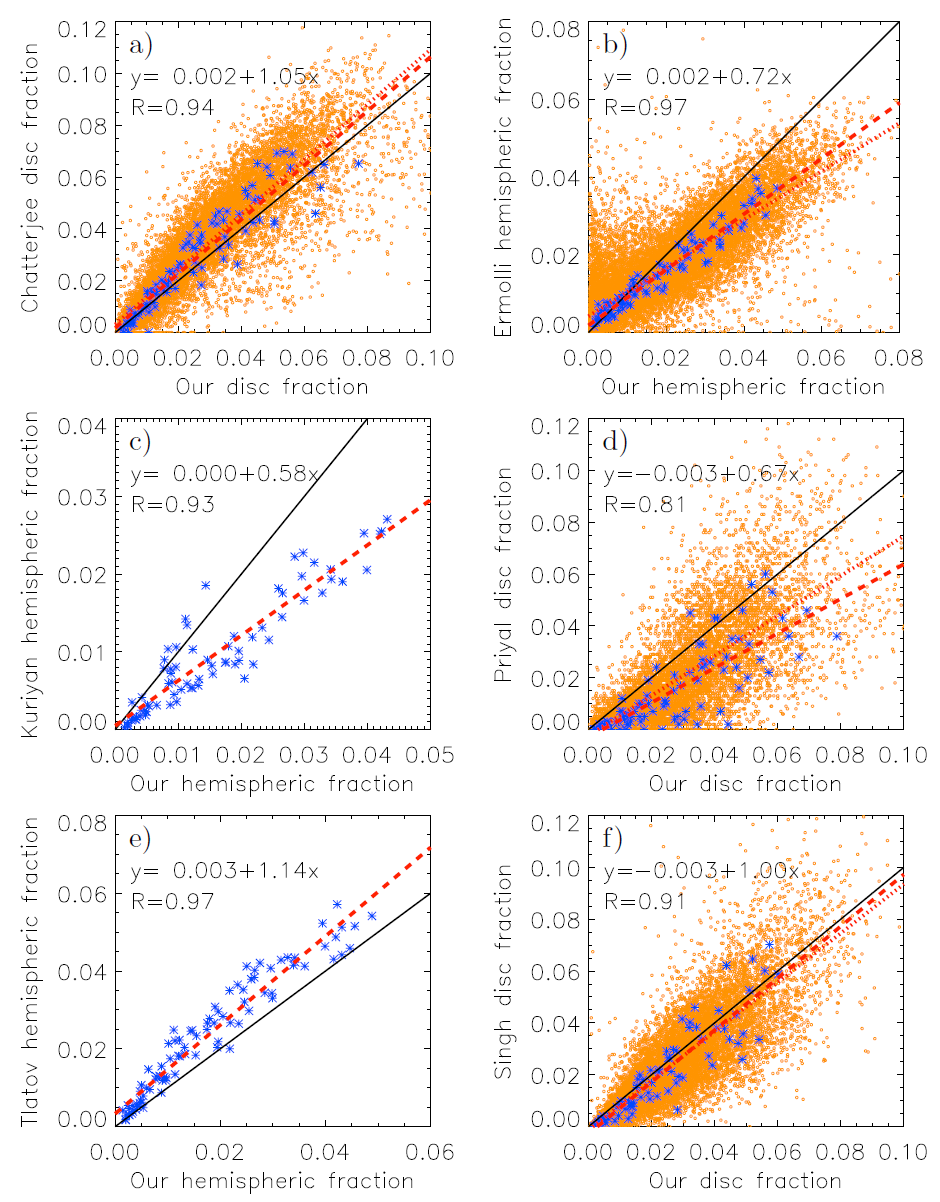}
	\caption{Plage areas presented in the literature ($y$-axis) versus the ones derived here from DS2 ($x$-axis): (\textbf{a}) \citet[][]{chatterjee_butterfly_2016} from DS2; (\textbf{b}) \citet[][]{ermolli_comparison_2009} from DS1;
 (\textbf{c}) \citet{kuriyan_long-term_1983} from the actual photographs;		(\textbf{d}) \citet[][]{priyal_long-term_2017} from DS2;
		(\textbf{e}) \citet[][]{tlatov_new_2009} from DS1; (\textbf{f}) \citet[][]{singh_variations_2018} from DS2. Blue asterisks (orange dots) show the annual (daily) values. The solid-black lines have a slope of unity. The dashed (dotted) red lines are linear fits to the annual (daily) data. Also shown are the corresponding parameters of the linear fits to the annual values and the linear correlation coefficients of the annual values.}
		\label{fig:scatterplots}
\end{figure*}

Figure \ref{fig:scatterplots} shows scatter plots between the plage areas derived by us from the DS2 data and the various published series obtained from Kodaikanal observations. 
In particular we consider the series by \cite{kuriyan_long-term_1983} derived from the physical photographs, the series by \cite{ermolli_comparison_2009} and \cite{tlatov_new_2009} from the DS1 data, and the series by \cite{chatterjee_butterfly_2016}, \cite{priyal_long-term_2017}, and \cite{singh_variations_2018} from the DS2 data. 
Note that the series by \cite{kuriyan_long-term_1983} and \cite{tlatov_new_2009} are only available as annual values and most likely include different selections of observations. 
For all the other series, we consider only the days common with the data from DS2 we use here. 
Besides comparing the daily values, we also compute and compare annual median values. 
However, we note that potential errors in the dates and times of the different original archives as well as the copies of them used by the respective authors, affect our results. 
The scatter between the various series and ours is rather significant.
Table \ref{tab:agreement} lists the RMS difference and the Pearson coefficient between the various series when daily values are used.
The best agreement is found between our results and those by \cite{chatterjee_butterfly_2016} and \cite{singh_variations_2018}, although when annual values are used then the agreement is better with the series by \cite{ermolli_comparison_2009} and \cite{tlatov_new_2009}. However there are hints of a non-linearity between our plage areas and those by \cite{tlatov_new_2009}.
In \cite{chatzistergos_analysis_2019} we did a similar comparison, but considering the above series and our results derived from analysis of DS1. 
Comparing the scatter plots in Figure \ref{fig:scatterplots} and those of \cite{chatzistergos_analysis_2019} we find an improvement in the match between our series and that from \cite{chatterjee_butterfly_2016} with linear correlation factor increasing to 0.94 compared to 0.9 for the annual values. 
Our results for the plage areas from DS2 data show worse agreement with the series by \cite{ermolli_comparison_2009} and \cite{tlatov_new_2009} from analysis of DS1 data than our results from DS1 data. 
However, we also notice a worse agreement with our results from DS2 data and those by \cite{priyal_long-term_2017} and \cite{singh_variations_2018} from analysis of DS2 data than our results from DS1 data.

\begin{table*}
	\caption{Quantification of the agreement between different plage area series. The values above the diagonal are the RMS differences, while those below the diagonal are the Pearson coefficients, both computed for the common days. The number of common days is given within the brackets. The abbreviations CEA16, EEA09, PEA17, and SEA18 refer to the \cite{chatterjee_butterfly_2016}, \cite{ermolli_comparison_2009}, \cite{priyal_long-term_2017}, and \cite{singh_variations_2018} series, respectively.}
	\centering
	\begin{tabular}{lcccccc}
		\hline
&DS1&DS2&CEA16&EEA09&PEA17&SEA18\\
DS1&-&0.006 [18387]&0.013 [12657]&0.012 [18454]&0.012 [10992]&0.009 [14370]\\
DS2&0.955 [18387]&-&0.011 [15479]&0.015 [20385]&0.015 [13112]&0.010 [16357]\\
CEA16&0.886 [12657]&0.896 [15479]&-&0.020 [13492]&0.019 [12499]&0.014 [12940]\\
EEA09&0.879 [18454]&0.838 [20385]&0.845 [13492]&-&0.013 [11593]&0.013 [14511]\\
PEA17&0.827 [10992]&0.784 [13112]&0.777 [12499]&0.780 [11593]&-&0.011 [12063]\\
SEA18&0.885 [14370]&0.867 [16357]&0.857 [12940]&0.827 [14511]&0.854 [12063]&-\\
			\hline
	\end{tabular}
	\label{tab:agreement}
\end{table*}

We now discuss various factors that are responsible for part of the differences.
Such factors are the definition of the centre coordinates and radius, the fraction of the disc used for the normalisation of the plage areas, as well as the processing techniques including the photometric calibration.

For example, different definitions of radius and centre coordinates affect the results of all series. 
\cite{priyal_long_2014} defined the radius and centre coordinates for the DS2 by manually selecting three points at the limb. 
This information was not stored, but rather was used to crop and centre the images. 
These cropped images were then used by \cite{chatterjee_butterfly_2016}, \cite{priyal_long-term_2017}, and \cite{singh_variations_2018} to derive their plage areas. 
\cite{ermolli_comparison_2009} and \cite{tlatov_new_2009} defined the radius and centre coordinates for DS1 independently.
\cite{chatzistergos_analysis_2019} used the radius estimates for DS1 made by \cite{ermolli_comparison_2009}, but we corrected a few errors.
Here for DS2 we determined the radius from scratch by using the method described by \cite{chatzistergos_analysis_2019}.

\begin{figure}
	\centering 
		\begin{overpic}[width=1\textwidth]{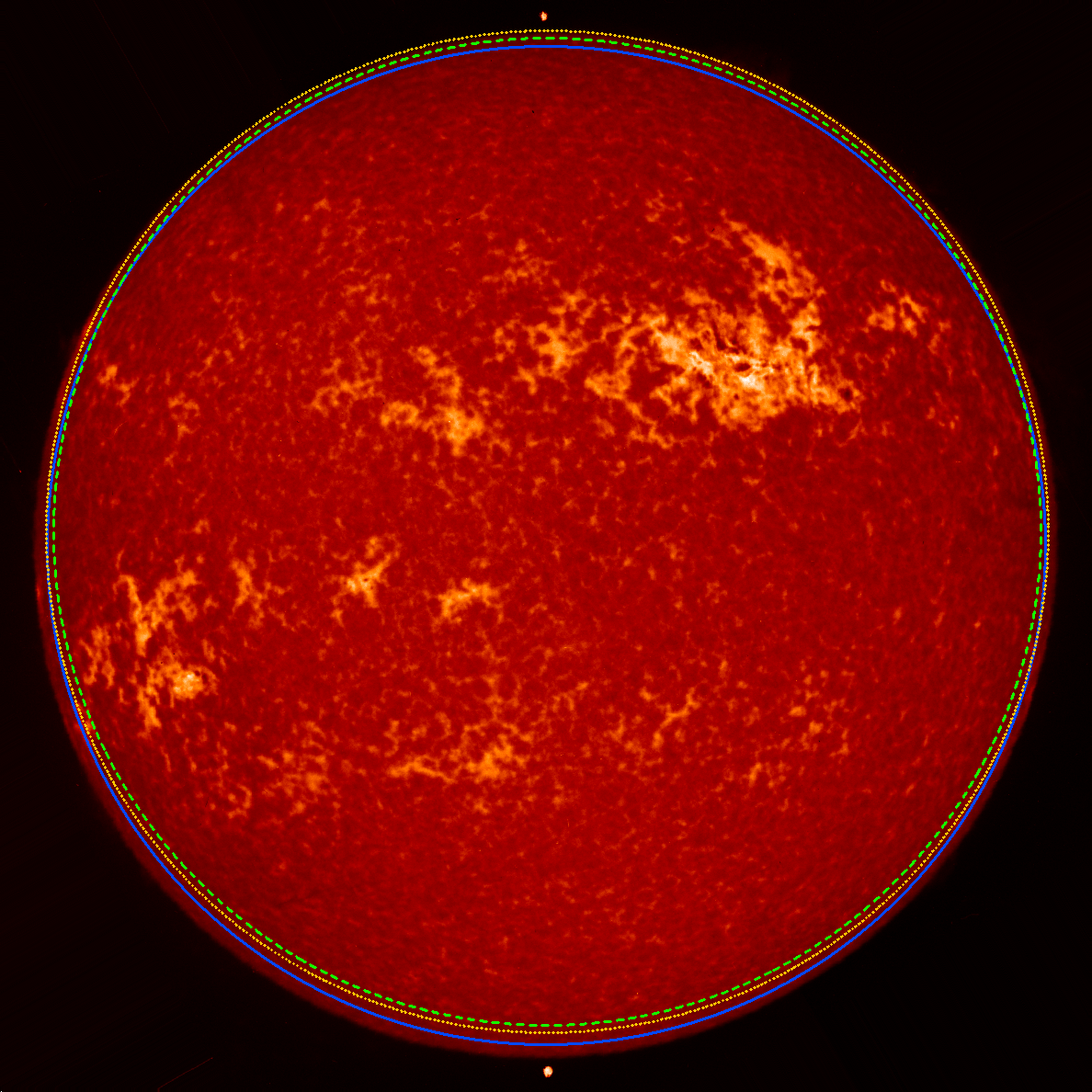} \end{overpic}			
	\caption{Raw Kodaikanal observation taken on 20 January 1938. Circles enclose the areas considered by \citet[][dashed green]{chatterjee_butterfly_2016}, \citet[][dashed green]{priyal_long-term_2017}, \citet[][dotted yellow]{priyal_long_2014}, and in this work (solid blue).}
	\label{fig:comparisonchatterjee}
 \end{figure}

Studies in the literature also differ in the fraction of the disc area that was used to normalise the identified plage areas.
For our study we used the disc area reaching out to 0.98$R$ (i.e., 96\,\% of the total area), 
while \cite{ermolli_comparison_2009}, \cite{chatterjee_butterfly_2016}, and \cite{priyal_long-term_2017} used the disc up to 0.97$R$ (94\,\% of the total area) \cite{priyal_long_2014} used the disc up to 0.985$R$ (97\,\% of the total area). 
Figure \ref{fig:comparisonchatterjee} shows an example observation from Kodaikanal where the regions considered in the various studies have been marked. Notice that the various studies defined the centre of the solar disc differently.
To get an error estimate for using a slightly different normalising area, we repeated the segmentation of DS1 data by considering the area of the disc up to 0.97$R$.
We found relative differences in the derived plage areas when considering the disc up to 0.97$R$ to 0.98$R$ that are $0.01\pm0.04$. 

\begin{figure}
	\centering 
	\includegraphics[width=1\linewidth]{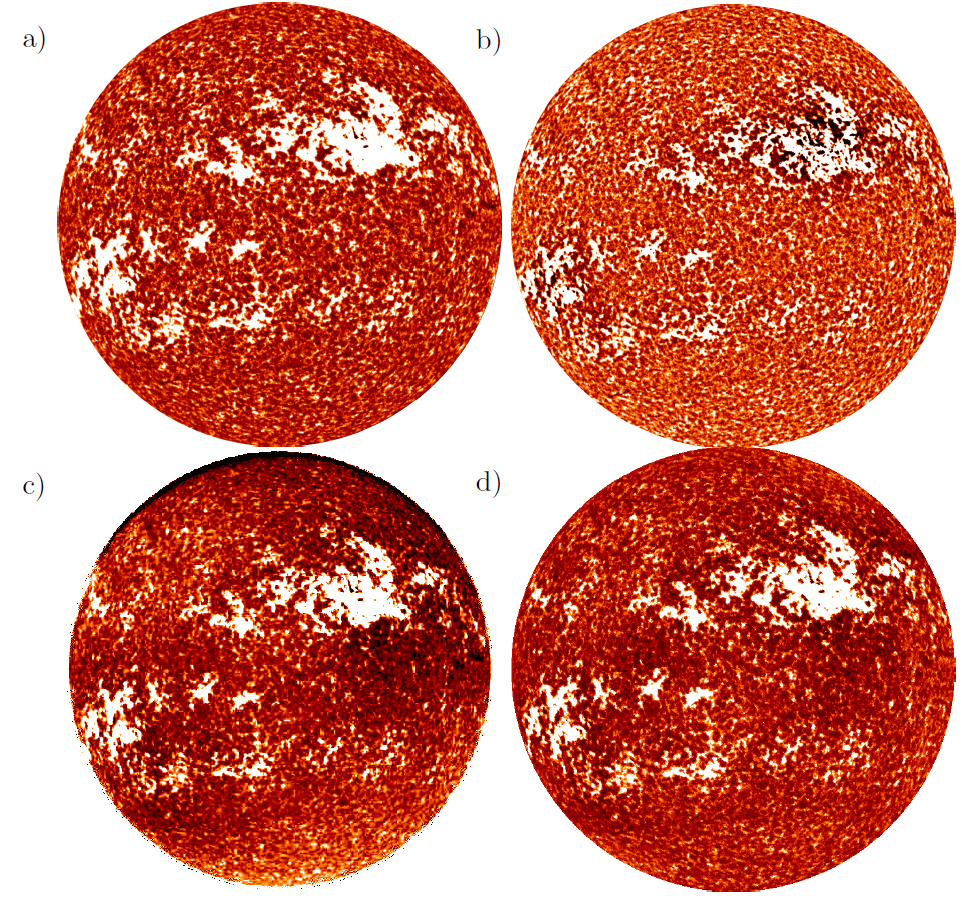}
	\caption{Calibrated images of Kodaikanal observation taken on 20 January 1938 (shown in Figure \ref{fig:comparisonchatterjee}): (\textbf{a}) with our method, and with the methods by (\textbf{b}) \citet[][]{chatterjee_butterfly_2016}, (\textbf{c}) \citet[][]{priyal_long_2014}, and (\textbf{d}) \citet[][]{ermolli_comparison_2009}. The images processed with our method and by \citet{ermolli_comparison_2009} are given as contrast values and are saturated at the same level [-0.02,0.02]. The image processed by \cite{chatterjee_butterfly_2016} is in arbitrary units, while the one by \cite{priyal_long_2014} was provided as a JPG image. Therefore, these images are saturated such that the plage regions visually appear similar to the saturated image with our method. }
	\label{fig:comparisonchatterjee2} \end{figure}

Processing artefacts contribute more to the systematic differences in the various results presented in the literature.  
For example, Figure \ref{fig:comparisonchatterjee2} shows the observation from Figure \ref{fig:comparisonchatterjee} calibrated with our method and with those used by \cite{ermolli_comparison_2009,priyal_long_2014,chatterjee_butterfly_2016}.
The QS regions in the image processed with our method are more uniform compared to the others. 
The images processed by \cite{ermolli_comparison_2009} and \cite{priyal_long_2014} show remaining large-scale inhomogeneities that can affect the plage area determination with these methods. 
In the image processed by \cite{chatterjee_butterfly_2016} the large-scale inhomogeneities have been accounted for, but the contrast of the plage regions has been suppressed, causing the immediately surrounding areas of large plage regions to become much darker. Thus a smaller part of the large plage areas will be considered as plage, while some network elements might be counted as plage.

\section{Conclusions}
\label{sec:conclusions}
The photographic archive of full-disc Ca {\sc II} K observations of the Kodaikanal observatory is a valuable source of information on past solar activity.
We have compared the two more recent digitisations of this archive to understand if there are any differences in the results and if these differences are responsible for the partly conflicting results presented in the literature. 
For this, we have processed images from the two digitisations consistently and applying the same technique.
Thus, we have applied the methods developed and tested by \cite{chatzistergos_analysis_2018} to the 16-bit data series as well.
We applied the same processing on modern CCD-based data from Rome/PSPT.

The plage areas derived from DS1 and DS2 and their variation with time are rather similar to each other. Many of the issues previously reported about the varying quality of the Kodaikanal data are found to apply to the data from the new digitisation too, implying that they are intrinsic characteristics of the physical archive.
These are an increase with time in the disc eccentricity coupled with a worsening spatial resolution, growing large-scale inhomogeneities, and change of the QS CLV with time.
However, we found the quality of the DS1 data after 1990 to deteriorate more than that of the DS2 data. This can introduce significant errors when cross-calibrating plage area series from different archives or for irradiance reconstructions.
Furthermore, both digitisations of Kodaikanal data seem to suffer from errors in the meta-data, especially concerning the observational date and time of the plates. 
This issue plagues all analyses from these data.

We find a good match between the plage areas derived from the Kodaikanal archive and those from the Rome/PSPT, with the Kodaikanal areas being slightly lower. 
This is in agreement with a drift in the quality of the Kodaikanal data and considering that Rome/PSPT has a nominal bandwidth which is five times broader than that of the Kodaikanal observatory.
The plage areas presented in the literature from the various digitisations of Kodaikanal data show significant differences. We suggest that the diverse employed methods to calibrate the data as well as the different definitions of the recorded solar radius are the main reasons for the discrepancies among the various published results. 

Overall, we found the new digitisation of the Kodaikanal archive to offer an improvement in the image quality over the 8-bit series. It also has more than doubled the available images.
With over 48,000 images, the new Kodaikanal series is possibly the richest currently available Ca {\sc II} K archive, so that it has a great potential to improve our understanding of solar activity, especially when the various issues affecting the series have been properly addressed.

\begin{acks}
The authors thank the Kodaikanal and the Rome Solar Groups. 
The newly digitized Kodaikanal data as presented here (DS2) are available at \url{kso.iiap.res.in/}. 
D. Banerjee. and I. Ermolli thank the International Space Science Institute (Bern, Switzerland) for supporting the International Team 420 "Reconstructing solar and heliospheric magnetic field evolution over the past century". 
T. Chatzistergos acknowledges a postgraduate fellowship of the International Max Planck Research School on Physical Processes in the Solar System and Beyond.
This work was supported by grants COST Action ES1005 "TOSCA", FP7 SOLID, 01LG1909C from the German Federal Ministry of Education and Research, and by the BK21 plus program through the National Research Foundation (NRF) funded by the Ministry of Education of Korea. 
This research has received funding from the European Union's Horizon 2020 research and innovation program under grant agreement No 824135 (SOLARNET).
We thank the anonymous referee for the valuable comments which helped to improve this manuscript.
This research has made use of NASA's Astrophysics Data System.
\end{acks}

\textbf{Disclosure of Potential Conflicts of Interest}\\
The authors declare that they have no conflicts of interest.

%
%
 \bibliographystyle{spr-mp-sola}
 \bibliography{_biblio}  
%
%
%
%

\end{article} 
\end{document}